\documentclass{article}
\begin{document}
\title{Twisted K--Theory in $g>1$ from D--Branes}

\author{Jos\'e M. Isidro\\
Department of Theoretical Physics,\\ 
1 Keble Road, 
Oxford OX1 3NP, UK.\\ 
{\tt isidro@thphys.ox.ac.uk}}
\maketitle

\begin{abstract}
We study the wrapping of $N$ type IIB D$p$--branes on a compact Riemann 
surface $\Sigma$ in genus $g>1$ by means of the Sen--Witten construction, 
as a superposition of $N'$ type IIB D$p'$--brane/antibrane pairs, 
with $p'>p$. A background Neveu--Schwarz field $B$ deforms the commutative 
$C^{\star}$--algebra of functions on $\Sigma$ to a noncommutative $C^{\star}$--algebra.
Our construction provides an explicit example of the $N'\to\infty$ limit 
advocated by Bouwknegt--Mathai and Witten 
in order to deal with twisted K--theory. We provide the necessary 
elements to formulate M(atrix) theory on this new $C^{\star}$--algebra, by
explicitly constructing a family of projective $C^{\star}$--modules admitting 
constant--curvature connections. This allows us to define the $g>1$ analogue
of the BPS spectrum of states in $g=1$, by means of Donaldson's formulation 
of the Narasimhan--Seshadri theorem. 



Preprint no. OUTP-01-58P.

\end{abstract}

\tableofcontents

\section{Introduction}\label{intro}  
  
\subsection{Setting}\label{set}  
  
The fact that D--branes carry vector bundles has allowed to  
interpret D--brane  charges and fields as classes in the K--theory
of spacetime, rather than as integer cohomology classes 
\cite{MINASIANMOORE}--\cite{MALDAMOOSEI}. This identification  
has led  to a better understanding of the spectrum of D--branes, 
in particular of  stable, nonsupersymmetric D--branes. Such non--BPS  
branes can often be  understood as bound states of a  
brane--antibrane system with tachyon condensation \cite{SEN}. 
 
It has been proposed \cite{WITMICHIGAN, BOUWMATHAI} that the K--theory 
analysis of a superposition of $N'$ type IIB D$p'$--brane/antibrane pairs 
is best performed in the limit $N'\rightarrow\infty$. This limit allows 
for the possibility of considering a nontorsion class for the field strength 
$H=dB$ of the Neveu--Schwarz field $B$.   
  
Along a related line, M(atrix) theory \cite{MATRIX, JAP}  
as a model for M--theory has been compactified toroidally in  
\cite{TAYLOR}. By turning on a background $B$--field one can deform 
this compactifation to a compactification on the noncommutative 
torus \cite{CDS, HULL}. The effective gauge theory on the D--branes
then becomes noncommutative \cite{SWn}.   
  
\subsection{Aim}\label{aim}  
  
In this paper we combine the three lines named above. The aim is to provide 
a physical interpretation for the $C^{\star}$--algebra 
constructed abstractly in \cite{PROCEEDINGS}. The strategy is as follows.

We first wrap $N$ type IIB D$p$--branes on a manifold $\Sigma\times Y$,
where $\Sigma$ is compact Riemann surface with genus $g>1$ and $Y$ is an 
auxiliary spacetime manifold to be specified presently.
(With more generality, one could consider a nontrivial bundle over $\Sigma$
instead of $\Sigma\times Y$.)
Following \cite{SEN}, each one of the $N$ wrapped 
D$p$--branes can be viewed as a superposition  of $N'$ D$p'$--brane/antibrane  pairs, 
with an odd value of $p'>p$. When wrapping a single type IIB D$p$--brane on a manifold 
of codimension $2k$, a minimum of $N'=2^{k-1}$ type IIB D$p'$--brane/antibrane pairs 
are needed \cite{WITTENDK}. Eventually passing to the limit $N\rightarrow\infty$ 
will also enforce $N'\rightarrow\infty$ and bring us into the stable range of K--theory. 
Simultaneously we turn on a background $B$--field across $\Sigma$. 

On the other hand, this system possesses a dual description in type IIA  
string theory or, more precisely, in 11--dimensional M--theory as described  
by M(atrix) theory. In this dual setting, a D$p$--brane is compactified on
$p$ copies of $S^1$, next T--dualised along all $p$ circles, and
finally decompactified into a type IIA D0--brane. The limit  $N\rightarrow\infty$ 
required by M(atrix) theory has a natural counterpart in the dual type IIB 
description: it arises from the requirement of allowing for the possibility 
that the background field strength $H$ be a nontorsion class.  
Our model provides an explicit realisation, in string theory terms,  
of the twisted K--theory described abstractly by Bouwkegt and Mathai 
in \cite{BOUWMATHAI}, and advocated by Witten in a similar 
K--theoretic setting \cite{WITMICHIGAN}. 

{}From this M(atrix) theory description of the wrapped D$p$--branes,  
the connection with noncommutative geometry \cite{NCG} is now immediate:
the background $B$--field deforms the commutative $C^{\star}$--algebra of 
functions on $\Sigma$ to a noncommutative $C^{\star}$--algebra.
  
\subsection{Outline}\label{outline}  

This paper is organised as follows.  As a preparation for $g>1$,
section \ref{morita} reviews the noncommutative torus from the 
standpoint of the Heisenberg algebra. The latter can be interpreted 
as a {\it central--curvature condition} on a projective module over 
the noncommutative torus \cite{SCHWARZ, KONECHNYSCHWARZ}.
({\it Central} means that, as an endomorphism of the projective module, 
the curvature is proportional to the identity. By abuse of terminology, 
we will call equation (\ref{central}) below a {\it constant--curvature condition}, 
rather than a central--curvature condition).

The constant--curvature condition has a natural extension to $g>1$ 
in the theory of stable, holomorphic vector bundles over a Riemann surface 
$\Sigma$, together with Donaldson's version \cite{DONALDSON} 
of the Narasimhan--Seshadri theorem \cite{NASE}. 
The latter provides the right mathematical description of the twisted 
gauge bundles arising on the stack of coincident branes required by the 
Sen--Witten construction of non--BPS branes. Indeed such bundles 
can be characterised as admitting a constant--curvature connection.  
These points are summarised in section \ref{nstheorema}.  

Section \ref{gimmeanameforthis} presents this new $C^{\star}$--algebra. 
We recall from \cite{PROCEEDINGS} the definition of its generators and 
of the trace required to write down the M(atrix) theory action, 
and explicitly construct the corresponding projective $C^{\star}$--modules.  

Wrapping a D$p$--brane on a closed, $(p+1)$--dimensional submanifold of spacetime 
is possible only when the condition of cancellation of global worldsheet anomalies 
is satisfied \cite{WITTENDK, BARYONS, FREEDWITTEN, KAPUSTIN}. This point is dealt 
with in section \ref{ann}. In particular, this analysis fixes the  
dimensionality of the D$p$--branes to be $p\geq 3$; this bound will be 
later refined by cohomological arguments in section \ref{manifoldy}.

Section \ref{background} presents, following \cite{BOUWMATHAI, KAPUSTIN},
the necessary formalism about the background field strength, oriented towards 
the limit $N\to\infty$ that will be taken in section \ref{wrapping}. 

In section \ref{wrapping} we first describe the setup in type IIB string theory terms. 
Next we pass, through a duality transformation, to an equivalent M(atrix) theory 
description of the $N$ D$p$--branes wrapped on $\Sigma$. As $N\rightarrow\infty$, 
so too must the 't Hooft magnetic flux $M$ go to infinity, in a certain sense to be 
specified presently. We will analyse this {\it double scaling limit} in detail; 
our $C^{\star}$--algebra of \cite{PROCEEDINGS} is precisely the double scaling limit 
of the Narasimhan--Seshadri representations of the Fuchsian group $\Gamma$ 
uniformising the Riemann surface $\Sigma$ in $g>1$. 

In section \ref{morgplus} we use Donaldson's theorem to identify the $g>1$ 
analogues of BPS states on the noncommutative torus, by explicitly identifying 
constant--curvature connections on the projective $C^{\star}$--modules 
constructed in section \ref{gimmeanameforthis}. 

Finally, section \ref{outlook} presents some conclusions and perspectives. 

\section{BPS spectra in $g=1$ from the Stone--von Neumann theorem}\label{morita}  
   
\subsection{The constant--curvature condition}\label{ccurvcon}  
  
Let us set the fermions of the M(atrix) theory action to zero, 
and consider a state determined by the condition that a connection on a 
projective module over the noncommutative torus $T_{\theta}^{2}$ 
have constant field strength,  
\begin{equation}  
F_{jk}=\omega_{jk}{\bf  I},  
\label{consteffe}
\end{equation}  
{\it i.e.}, the curvature must be proportional to the identity endomorphism.  
Above, $\omega_{jk}$ is a constant 2--form over the Lie algebra of  
derivations of $T_{\theta}^2$. In the presence of supersymmetry such field   
configurations give rise to BPS states \cite{CDS, SCHWARZ}, with an amount of preserved   
supersymmetry given by the dimension of the space of spinors $\epsilon$, 
$\epsilon'$ that solve the equation  
\begin{equation}  
\epsilon\Gamma^{jk}\, F_{jk} + \epsilon' {\bf  I}=0,  
\label{spinor}
\end{equation}  
where $\Gamma^{jk}=[\Gamma^j, \Gamma^k]$ is a commutator of Dirac matrices. 
In the absence of supersymmetry, as will be the case in $g>1$, 
condition (\ref{consteffe}) is the closest analogue 
of the equation (\ref{spinor}) defining a BPS state.  
  
In \cite{KONECHNYSCHWARZ} it has been argued that the complete set  
of equations specifying a projective module over the torus $T_{\theta}^{2}$, 
together with a constant curvature connection on it, is given by  
\begin{equation}  
{\cal U}_j\,{\cal U}_k=e^{2\pi i\theta_{jk}}{\cal U}_k\,{\cal U}_j,\qquad  
\left[\nabla_j, {\cal U}_k\right]=\delta_{jk}\,{\cal U}_k,\qquad  
\left[\nabla_j,\nabla_k\right]=iF_{jk}\,{\bf  I},  
\label{konsch}
\end{equation}  
where $j,k=1,2$. These equations can be solved by first representing  
the Heisenberg algebra  
$\left[\nabla_j,\nabla_k\right]=iF_{jk}\,{\bf  I}$, through the  
Stone--von Neumann theorem, on the Hilbert space $L^2({\bf R})$. 
The hermiticity of this representation ensures the unitarity of the generators 
\begin{equation}
{\cal U}_j={\rm exp}\,(iF^{-1}_{jk}\nabla_k).
\label{genecalu}
\end{equation}
Next we tensor the latter with an $N\times N$ dimensional representation of  
't Hooft's matrices $u_j$  
\begin{equation}  
u_ju_k=e^{2\pi i M_{jk}/N}u_ku_j,\qquad M_{jk}\in{\bf Z},  
\label{thoftmat}
\end{equation}  
acting on the space ${\bf C}^N$. The complete projective module over 
$T_{\theta}^{2}$ is given by
\begin{equation}
E_{NM}=L^2({\bf R})\otimes {\bf C}^N_{(M)},
\label{totamod}
\end{equation}
where the notation ${\bf C}^N_{(M)}$ makes reference to the magnetic flux 
$M=M_{12}$.  The total generators 
\begin{equation}
{\cal U}_j\otimes u_j, \qquad j=1,2,
\label{turuturu}
\end{equation}
satisfy the algebra of $T_{\theta}^{2}$ with a total deformation 
parameter
\begin{equation}
\theta_{jk}=-{1\over 2\pi } F_{jk} + {1\over N}M_{jk}.
\label{totdefpara}
\end{equation}
The fact that $F_{jk}\,{\bf  I}$ is a c--number allows one to compute 
the deformation parameter by a simple application of
the Baker--Campbell--Hausdorff formula.  
  
\subsection{Moduli space of constant--curvature connections}\label{modspace}  
  
The notion of a moduli space ${\cal M}^{(g=1)}$ of constant--curvature 
connections in $g=1$ appears naturally in the above picture \cite{KONECHNYSCHWARZ}. 
${\cal M}^{(g=1)}$ is the space of solutions to the first two equations 
of (\ref{konsch}), modulo gauge transformations. Modules possessing different 
Chern numbers are treated simultaneously in this approach. Fixing a Chern number 
corresponds to choosing a connected component of the total moduli space of solutions 
to equation (\ref{konsch}).  
  
The residual gauge transformations preserving the constant--curvature condition  
(\ref{konsch}) correspond to $N\times N$ unitary transformations acting on the  
${\bf C}^N$ factor of the  module $E_{NM}$. Hence the moduli space of  
constant--curvature connections on a module with fixed integer values of $N$  
and $M_{jk}$ can be described as a space of inequivalent representations of  
the matrix algebra (\ref{thoftmat}). The latter in fact admits a continuum of  
inequivalent representations. In order to identify it, we first consider the  
commutative torus $\hat T^{2}$ that is dual to the original commutative torus 
$T^{2}$. Then the space of irreducible representations 
of (\ref{thoftmat}) is  described by means of $2$ complex numbers $\lambda_i$ 
with unit modulus, modulo a certain residual gauge symmetry. Let $E_{\Lambda}$, 
$\Lambda=(\lambda_1, \lambda_{2})$, denote the corresponding irreducible 
representations, and assume that ${\bf C}^N$ decomposes  
as ${\bf C}^N=\oplus_{l=1}^r E_{\Lambda_l}$. The residual gauge  
symmetry acts by permutation on the $r$ summands as the permutation group  
$S_r$, and the moduli space ${\cal M}^{(g=1)}$ is $\hat T^{2}/S_r$.  

\section{The Narasimhan--Seshadri theorem}\label{nstheorema} 
  
\subsection{Statement of the theorem}\label{useful}  
  
Let $\Gamma$ denote the Fuchsian group uniformising a compact 
Riemann surface $\Sigma$ with genus $g>1$ and without boundary. 
We now summarise some facts about projective, unitary representations 
of $\Gamma$ and the theory of holomorphic vector bundles over $\Sigma$ 
\cite{NASE} (for more extensive treatments see \cite{KOBAYASHI, ABZT}).

Let ${\cal  E}\rightarrow \Sigma$  be a holomorphic vector bundle over $\Sigma$ 
of rank $N$ and degree, {\it i.e.} first Chern class, $M$. The bundle ${\cal  E}$ 
is called {\it stable} if the ratio  
\begin{equation}  
\mu({\cal  E})={M\over N}  
\label{ratio}  
\end{equation}  
satisfies the inequality  
$\mu({\cal  E}')<\mu({\cal  E})$ for every proper holomorphic subbundle   
${\cal  E}'\subset {\cal  E}$. We may take $-N<M\leq 0$, as this may always 
be arranged by tensor multiplication with a line bundle without losing stability. 

Denote by $\gamma_j$, $j=1,\ldots, 2g$, the generators of $\Gamma$. We have
\begin{equation}  
\prod_{j=1}^g \left(\gamma_{2j-1}\,\gamma_{2j}\,
\gamma_{2j-1}^{-1}\, \gamma_{2j}^{-1}\right)={\bf I}.  
\label{genex}  
\end{equation}
For the purposes of this section we will temporarily assume that  
$\Gamma$ contains a unique primitive elliptic element $\gamma_0$ of 
order $N$, {\it i.e.} $\gamma_0^N={{\bf  I}}$, with fixed point $z_0\in{{\bf  
H}}$ that projects to $x_0\in\Sigma$. Now let $\rho:\Gamma\to U(N)$ 
be an irreducible unitary representation. It is said  {\it admissible} if  
\begin{equation}  
\rho(\gamma_0)=e^{-2\pi i M/N}{{\bf  I}}.  
\label{ammissibile}
\end{equation}  
Putting the elliptic element on the right--hand side, and denoting  
$\rho(\gamma_j)$ by  $u_j$, an admissible representation satisfies  
\begin{equation}  
\prod_{j=1}^g \left(u_{2j-1}\,u_{2j}\,u_{2j-1}^{-1}\, u_{2j}^{-1}\right)=  
e^{2\pi iM/N}{{\bf  I}}.  
\label{nsmatrices}
\end{equation}  
The $u_j$ are the $g>1$ generalisation of 't Hooft's matrices (\ref{thoftmat}).  
  
On the trivial bundle ${{\bf H}}\times {{\bf C}}^N\rightarrow {{\bf  
H}}$ there is an action of $\Gamma$:  
$(z, v)\mapsto(\gamma z, \rho(\gamma)v)$.  This defines the quotient  
\begin{equation}  
{{\bf H}}\times {{\bf C}}^N/\Gamma\rightarrow {{\bf H}}/\Gamma\cong\Sigma.  
\label{trivialquot}  
\end{equation}  
Any admissible representation determines a holomorphic vector bundle  
${\cal  E}_{\rho}\rightarrow \Sigma$ of rank $N$ and degree $M$. When $M=0$,  
${\cal  E}_{\rho}$ is simply the  quotient bundle (\ref{trivialquot}).  
The Narasimhan--Seshadri theorem  now states that a holomorphic vector bundle 
${\cal  E}\rightarrow\Sigma$ of rank $N$ 
and degree  $M$  is stable if and only if it is isomorphic to a  bundle 
${\cal  E}_{\rho}$, where $\rho$  is an admissible representation of $\Gamma$. 
Moreover, the bundles ${\cal  E}_{\rho_1}$ and ${\cal  E}_{\rho_2}$ are isomorphic  
if and only if the representations $\rho_1$ and $\rho_2$ are equivalent.

Next consider the adjoint representation of $\Gamma$ on ${\rm End}\,{{\bf C}}^N$,  
\begin{equation}  
{\rm Ad}\,\rho (\gamma) Z =\rho (\gamma) Z \rho^{-1}(\gamma),  
\label{aggiunta}  
\end{equation}  
where $Z\in{\rm End}\,{{\bf C}}^N$ is understood as an $N\times N$ matrix.  Let us  
also consider the trivial bundle ${{\bf H}}\times{\rm End}\,{{\bf C}}^N\rightarrow  
{{\bf H}}$. There is an action of $\Gamma$:  
$(z,Z)\mapsto(\gamma z, {\rm Ad}\,\rho(\gamma) Z)$ that defines the quotient bundle  
\begin{equation}  
{{\bf H}}\times {\rm End}\, {{\bf C}}^N/\Gamma\rightarrow  
{{\bf H}}/ \Gamma\cong\Sigma.  
\label{quotient2}
\end{equation}  
When ${\cal  E}$ is stable, the bundle of endomorphisms ${\rm End}\,  
{\cal  E}\rightarrow \Sigma$ is isomorphic to the quotient bundle  
(\ref{quotient2}).  
  
\subsection{Donaldson's approach to stability of vector  
bundles}\label{donapp}  
  
A differential--geometric approach to stability has been given by Donaldson  
\cite{DONALDSON}. Fix a Hermitean metric on $\Sigma$, for example the Poincar\'e  
metric, normalised so that the area of $\Sigma$ equals 1. Let us denote by  
$\omega$ its associated (1,1)--form. Then a holomorphic vector bundle is stable 
if and only if it admits a metric connection $\nabla_D$ with constant curvature  
\begin{equation}  
F_D=-2\pi i \mu({\cal E}) \,\omega \,{{\bf  I}};
\label{central}
\end{equation} 
such a connection $\nabla_D$ is unique. As done for BPS states in $g=1$ 
\cite{SCHWARZ}, in section \ref{morgplus} we will use the constant--curvature 
condition (\ref{central}) to characterise BPS--like states in $g>1$.  

\section{Infinite--dimensional projective representations of the Fuchsian 
group $\Gamma$}\label{gimmeanameforthis}  
 
In order to study M(atrix) theory in $g>1$ the following elements are needed: 
a knowledge of the $C^{\star}$--algebra, a trace, and the projective $C^{\star}$--modules. 

\subsection{Definition of the $C^\star$--algebra $C^{\star}(\Gamma, \theta)$}\label{defcstar} 

Let us recall from \cite{PROCEEDINGS} the construction of the operators 
${\cal U}_j=\rho_b(\gamma_j)$ that provide a projectively unitary 
representation $\rho_b$ of the Fuchsian group $\Gamma$.
We first pick a fundamental domain ${\cal F}_z$ for the 
Fuchsian group $\Gamma$ uniformising $\Sigma$, with basepoint  
$z\in{\bf H}$, in order to have a tessellation $T({\bf H})$ of ${\bf H}$.   
On the Hilbert space $L^2({\bf H})$ one defines,   
for every value of the Fuchsian index $j=1,\ldots, 2g$, 
\begin{equation}  
{\cal U}_j^{(z)}={\rm exp}\left(ib\int_z^{\gamma_j z} A\right)  
\prod_{\alpha=-1}^1{\rm  
exp}\left[\lambda_{\alpha}^{(j)}(L_{\alpha}+\bar L_{\alpha})\right].
\label{genu}
\end{equation}  
Above, the $L_{\alpha}$, $\bar L_{\alpha}$ are the standard 
${\rm sl}_2({\bf R})$ differential generators $z^{\alpha +1}\partial_z$, 
$\bar z^{\alpha +1}\partial_{\bar z}$, $A={\rm d}{\rm Re}(z)/{\rm Im} (z)$ 
is a gauge field on ${\bf H}$, 
the $\lambda_{\alpha}^{(j)}$ are a set of numerical parameters specifying  
a complex structure on $\Sigma$, and $b$ is an arbitrary real parameter.
One can prove that the ${\cal U}_j^{(z)}$ are unitary 
and satisfy 
\begin{equation}  
\prod_{j=1}^g\left({\cal U}_{2j-1}\,{\cal U}_{2j}\,{\cal U}_{2j-1}^{-1}\,  
{\cal U}_{2j}^{-1}\right)=e^{-2\pi i\theta_b}{{\bf  I}},  
\label{inoxnni}
\end{equation}  
with $\theta_b$ a real parameter that is independent of the basepoint $z$ 
and is given by 
\begin{equation}
\theta_b=b\chi(\Sigma)=b(2-2g).
\label{valueoftheta}
\end{equation}
Consider the associative algebra with involution whose unitary generators 
are the ${\cal U}_j^{(z)}$ of equation (\ref{inoxnni}). 
It admits a faithful unitary representation on $L^2({\bf H})$. 
Taking the norm closure of this image \cite{NCG}, this algebra becomes 
a $C^\star$--algebra that we denote by $C^{\star}(\Gamma, \theta)$. 

\subsection{Definition of the trace}\label{deftrace} 
 
A trace can be defined by means of the following equivalent presentation of    
$C^{\star}(\Gamma, \theta)$ \cite{PROCEEDINGS}. Each $\gamma\neq {{\bf  I}}$   
in $\Gamma$ can be univocally expressed as a positive power of a primitive element   
$\tilde p\in\Gamma$, {\it primitive} meaning that it is not a positive power of  
any other element in $\Gamma$ \cite{MCKEAN}. Let ${\cal V}_{\tilde p}$ 
be the representative of $\tilde p$. Any ${\cal V} \in C^\star$ can  be written as  
\begin{equation}  
{\cal V}=\sum_{\tilde p\in\{prim\}}\sum_{n=0}^{\infty}c_n^{(\tilde p)}{\cal 
V}_{\tilde p}^n + c_0{{\bf  I}},  
\label{prim}  
\end{equation}  
for certain coefficients $c_n^{(\tilde p)}$, $c_0$. We now define a trace as  
\begin{equation} 
{\rm tr}\,{\cal V}=c_0.  
\label{traccia}
\end{equation} 
  
\subsection{Construction of projective $C^{\star}(\Gamma,\theta)$--modules 
$E_{NM}$}\label{construction}  
  
The Hilbert space $L^2({\bf H})$  becomes a right 
$C^{\star}(\Gamma, \theta)$--module under right multiplication of 
$\xi\in L^2({\bf H})$ with the ${\cal U}_j^{(z)}$. A $C^{\star}(\Gamma, 
\theta)$--valued inner product $\langle\,,\rangle$ on this module 
can be defined by summing over the Fuchsian indices, 
and over the vertices $z\in T({\bf H})$:  
\begin{equation}  
\langle \xi,\eta\rangle=\sum_{z\in T({\bf H})}\sum_{j=1}^{2g}\,  
(\xi,\eta\,{{\cal U}_j^{(z)}}^{\dagger})\,{\cal U}_j^{(z)},
\qquad \xi,\eta\in L^2({\bf H}).  
\label{inprod}
\end{equation}  
In equation (\ref{inprod}), $(\,,)$ denotes the Hermitean product 
on $L^2({\bf H})$ constructed with respect to the Poincar\'e metric 
on ${\bf H}$. Next we tensor the differential operators ${\cal U}_j^{(z)}$ 
with a set of Narasimhan--Seshadri matrices $u_j$. 
A projective $C^{\star}(\Gamma, \theta)$--module 
$E_{NM}$ is defined as the tensor product of $L^2({\bf H})$  
times the Narasimhan--Seshadri representation space ${\bf C}^N_{(M)}$ with degree $M$:
\begin{equation}
E_{NM}=L^2({\bf H})\otimes {\bf C}^N_{(M)}.
\label{totspace}
\end{equation}
The total generators on $E_{NM}$ are ${\cal U}_j^{(z)}\otimes u_j$, 
with the matrix part contributing a piece
\begin{equation}  
\langle \xi_N,\eta_N\rangle_N=\sum_{j=1}^{2g}\,  
(\xi_N,\eta_N\,u_j^{\dagger})\,u_j, \qquad   \xi_N,\eta_N\in {\bf C}^N 
\label{inprodmat}
\end{equation} 
to the scalar product on $E_{NM}$. In equation (\ref{inprodmat}), $(\,,)$ denotes 
the standard Hermitean product on ${\bf C}^N$. The total deformation parameter 
for the generators ${\cal U}_j^{(z)}\otimes u_j$ is then
\begin{equation}  
\theta_{\rm tot}=\theta_b-M/N.  
\label{totdef} 
\end{equation}
  
\section{The anomaly--cancellation condition}\label{ann}  
  
In type IIB superstring theory on a spacetime $X$, consider wrapping a   
D$p$--brane on a closed, $(p+1)$--dimensional submanifold $Q\subset X$. 
The analysis of global worldsheet anomalies for open superstrings attached 
to D$p$--branes has been performed in \cite{WITTENDK, FREEDWITTEN, KAPUSTIN}. 
Let us briefly summarise it.  
  
In the presence of a background Neveu-Schwarz 2--form $B$,  
a single D$p$--brane can be wrapped on a submanifold $Q\subset X$  
if and only if the normal bundle ${\cal N}$ of $Q$ satisfies the 
condition of cancellation of global anomalies for open superstrings 
ending on $Q$:   
\begin{equation}  
\beta_2(w_2({\cal N}))=[H]_Q.  
\label{normal}
\end{equation}  
Here $[H]$ is the integer cohomology class whose de Rham representative is   
$H=dB$, $[H]_Q$ denotes its restriction to $Q$, and  
$\beta_2(w_2({\cal N}))$   
is the image of the second Stiefel--Whitney class $w_2({\cal N}) 
\in H^2(Q,{\bf Z}_2)$   
under the Bockstein homomorphism $\beta_2:H^2(Q,{\bf Z}_2) 
\rightarrow H^3(Q, {\bf Z})$   
induced by the short exact sequence  
\begin{equation}  
0\rightarrow{\bf Z}\rightarrow{\bf Z}\rightarrow{\bf Z}_2\rightarrow 0.  
\label{shortuno}
\end{equation}  
Above, the second arrow is multiplication by 2, while the third arrow is   
reduction modulo 2.  
  
The wrapping of $N$ D$p$--branes on a submanifold $Q$ is governed by a   
generalisation of equation (\ref{normal}) that we describe next. When  
$[H]_Q=0$,  the $N$ D$p$--branes carry a $U(N)$ principal bundle while,  
for $[H]_Q\neq 0$, the D$p$--branes carry a principal $SU(N)/{\bf Z}_N$ 
bundle that cannot be lifted  to a $U(N)$ bundle.
Now the 't Hooft magnetic 2--form is a cohomology class   
$[f]\in H^2(Q,{\bf Z}_N)$. Consider the image of $[f]$ under  
the Bockstein homomorphism   
$\beta_N:H^2(Q,{\bf Z}_N)\rightarrow H^3(Q,{\bf Z})$ induced by  
the short exact sequence  
\begin{equation}  
0\rightarrow{\bf Z}\rightarrow{\bf Z}\rightarrow{\bf Z}_N\rightarrow 0,  
\label{shortdue}
\end{equation}  
where the second arrow is multiplication by $N$, while the  
third arrow is   
reduction modulo $N$. The image $\beta_N([f])\in 
 H^3(Q,{\bf Z})$ measures   
the obstruction to lifting an $SU(N)/{\bf Z}_N$ bundle  
to a $U(N)$ bundle.  
It turns out that global worldsheet anomalies for open superstrings   
ending on the D$p$--branes cancel if and only if
\begin{equation}  
\beta_N([f])+ \beta_2(w_2({\cal N}))=[H]_Q.  
\label{totalanomaly}
\end{equation}  
For the above condition to be nonempty it is required that $p\geq 3$.

\section{The background field strength}\label{background}  

\subsection{Local description of a twisted bundle}\label{local}  
  
An $SU(N)/{\bf Z}_N$ bundle without $U(N)$ structure has the following  
description in terms of transition functions. Take a good covering of $X$   
by open sets $W_i$, and denote by $su(N)$ the Lie algebra of $SU(N)/{\bf Z}_N$. 
A vector bundle associated with the principal $SU(N)/{\bf Z}_N$ bundle has sections 
$f_i:W_i\rightarrow su(N)$. Transition functions $g_{ij}:W_i\cap W_j\rightarrow U(N)$ 
are defined on double overlaps, such that   
\begin{equation}  
f_i=g_{ij}f_jg_{ij}^{-1}=g_{ij}f_jg_{ji},  
\label{trans}
\end{equation}  
while on triple overlaps $W_i\cap W_j\cap W_k$ the consistency condition  
\begin{equation}  
g_{ij}g_{jk}g_{ki}=h_{ijk}  
\label{consistency}
\end{equation}  
must be satisfied. Above, $h_{ijk}$ is an $N^{th}$ root of unity 
obeying the cocycle relation  
\begin{equation}  
h_{ijk}\,h_{ikl}=h_{jkl}\,h_{ijl}  
\label{cocyclei}
\end{equation}  
on quadruple overlaps. From here  
\begin{equation}  
\ln h_{ijk} + \ln h_{ikl} - \ln h_{jkl} - \ln h_{ijl}   
=2\pi i \kappa_{ijkl},  
\label{kappa}
\end{equation}  
where $\kappa_{ijkl}$ defines an element $\kappa\in H^3(X,{\bf Z})$ 
which is the obstruction to lifting the $SU(N)/{\bf Z}_N$ bundle to  
a $U(N)$ bundle.   
  
Therefore, in the presence of $[H]\neq 0$, gauge bundles on the  
branes are described by transition functions that obey  
equation (\ref{consistency}). The direct sum of two such twisted bundles 
obeys the same condition. Under the usual equivalence relation of K--theory, 
equivalence classes of twisted bundles define the twisted K--theory of $X$, 
denoted $K_{[H]}(X)$ \cite{WITTENDK}.  

\subsection{The Brauer group}\label{bru}

The background field strength $H$ determines a class in the  
\v Cech cohomology group $H^3(X,{\bf Z})$ \cite{BOTT}.
The latter decomposes as   
\begin{equation}  
H^3(X,{\bf Z})={\bf Z} \oplus \dots \oplus {\bf Z} \oplus {\bf Z}_{q_1}   
\oplus \ldots \oplus {\bf Z}_{q_s}.  
\label{decomposes}
\end{equation}  
The ${\bf Z}_q$ pieces are called {\it torsion subgroups}.  
Torsion classes  determine a subgroup of $H^3(X,{\bf Z})$, 
called the Brauer group of $X$ and denoted $Br(X)$. 
Next we give two different descriptions of the latter.   
One is in terms of finite--dimensional Azumaya algebras over $X$,   
the other one is through ${\cal K}$--bundles with structure  
group ${\rm Aut}({\cal K})$.  The link between these two descriptions 
of $Br(X)$ is explained in \cite{BOUWMATHAI}.   
 
\subsection{Azumaya algebras over $X$}\label{azumaya}  
  
Let $M_N({\bf C})$ denote the algebra of complex $N\times N$ matrices.   
Its automorphism group ${\rm Aut}(M_N({\bf C}))$ is  $PU(N)=SU(N)/{\bf Z}_N$,   
where $PU(N)=U(N)/U(1)$ denotes the projective unitary group on ${\bf C}^N$.  
  
An {\it Azumaya algebra over $X$} is a fibre bundle over $X$, whose typical 
fibre is the algebra $M_N({\bf C})$. Sections $f_i$ are $M_N({\bf C})$--valued 
and transition functions $g_{ij}$ are $PU(N)$--valued, in such a way that 
equations (\ref{trans})--(\ref{kappa}) above are satisfied.  
  
For any torsion class $[H]\in H^3(X,{\bf Z})$ there is a unique  
(equivalence class of) Azumaya algebras and the corresponding twisted 
K--theory, $K_{[H]}(X)$ \cite{KAPUSTIN, BOUWMATHAI}.  
  
\subsection{${\cal K}$--bundles over $X$}\label{kbundles}  
  
In the $C^{\star}$--norm topology, the limit \cite{DIXMIERL}
\begin{equation}
{\rm lim}_{N\to\infty}M_N({\bf C})={\cal K}
\label{limkappa}
\end{equation}
defines the $C^{\star}$--algebra ${\cal K}$ of compact operators on an infinite--dimensional, 
separable Hilbert space ${\cal H}$. Let $U({\cal H})$ denote 
the group of unitary operators on ${\cal H}$, and let $PU({\cal H})=U({\cal H})/U(1)$ 
be the projective unitary group on ${\cal H}$. By the same token we can set 
\begin{equation}
{\rm lim}_{N\to\infty} SU(N)/{\bf Z}_N=PU({\cal H}).
\label{sett}
\end{equation}   
Furthermore it holds that ${\rm Aut}({\cal K})=PU({\cal H})$.  
  
Let us consider a locally trivial bundle ${\cal  E}$ over $X$ with fibre   
${\cal K}$ and structure group ${\rm Aut}({\cal K})$. Such a bundle is also
determined by equations (\ref{trans})--(\ref{kappa}), where now the typical fibre 
is the algebra ${\cal K}$, hence sections $f_i$ are ${\cal K}$--valued   
and transition functions $g_{ij}$ are  $PU({\cal H})$--valued \cite{BOUWMATHAI}.  
   
\subsection{$H^3(X,{\bf Z})$ as parameter space for 
${\cal K}$--bundles}\label{parameter}  
  
Isomorphism classes of locally trivial bundles ${\cal  E}$ over $X$  
with fibre  ${\cal K}$ and structure group ${\rm Aut}({\cal K})$ are  
parametrised by $H^3(X, {\bf Z})$. With every {\it torsion class} in 
$H^3(X,{\bf Z})$ there is associated an isomorphism class of 
{\it projectively flat} bundles ${\cal E}$ with fibre ${\cal K}$ 
and structure group ${\rm Aut} ({\cal K})$ \cite{BOUWMATHAI}. 
Such bundles are given by a representation of $\pi_1(X)$ into 
${\rm Aut}({\cal K})$ \cite{KOBAYASHI}.   
  
The cohomology class in $H^3(X,{\bf Z})$ corresponding to a bundle 
${\cal  E}$ with fibre ${\cal K}$ and structure group ${\rm Aut}({\cal K})$  
is called the {\it Dixmier--Douady invariant} of ${\cal  E}$; it is denoted  
$\delta({\cal  E})$ \cite{DIXMIER}. In terms of transition functions,  
$\delta({\cal E})$ equals the cohomology class $\kappa$ given in equation 
(\ref{kappa}),  with the obvious replacements.  

\section{Wrapping D--branes on a $g>1$ Riemann surface}\label{wrapping} 

\subsection{The type IIB description}\label{braneantibrane}  

In what follows we take $Q$ to be a manifold of the form $\Sigma\times 
Y$, for some (as yet) unspecified manifold $Y$.  We want to wrap $N$ coincident
type IIB  D$p$--branes on $Q$. Forgetting about the manifold $Y$ for the moment, 
we will speak of $N$ coincident D$p$--branes wrapping $\Sigma$.

Now each one of those D$p$--branes, through the Sen--Witten construction 
\cite{WITTENDK, SEN}, can be thought of as a superposition of $N'=2^{k-1}$ 
D$p'$--brane/antibrane pairs on ${\bf R}^{p+1}$. Here $2k$ is the 
codimension of the D$p$--branes and $p'>p$. According 
to \cite{WITTENDK, SEN}, an appropriate choice for the tachyon field makes 
this superposition equivalent to a D$p$--brane wrapped on $\Sigma$. 
Eventually passing to the limit $N\to\infty$ will also enforce 
$N'\to\infty$, thus bringing us into the stable range of K--theory. 
This is in nice agreement with \cite{WITMICHIGAN, BOUWMATHAI}, 
where it has been proposed that the K--theory analysis of a superposition 
of $N'$ D$p'$--brane/antibrane pairs is best performed 
in the limit $N'\rightarrow\infty$.  

\subsection{The dual description: M(atrix) theory}\label{dual}  
 
Our system of $N$ D$p$--branes wrapped on $\Sigma$ has a dual description
that allows us to make contact with the setup of \cite{PROCEEDINGS}.
We first unwrap the D$p$--branes into flat space. Next we
compactify them along $p$ spatial coordinates, on $p$
copies of $S^1$. A further step is to apply a T--duality on all $p$ circles. 
Finally we decompactify them by sending their radii to infinity. 
The result is a system of $N$ D0--branes. So far the Riemann surface 
$\Sigma$ has played a spectator role. However, the  
$N$ D0--branes can be compactified on the original $\Sigma$. 
The resulting system is best understood in 11--dimensional M(atrix) 
theory compactified on the Riemann surface $\Sigma$,
as done in \cite{PROCEEDINGS}. For the rest of this paper we will
adhere to this dual picture. Then the limit 
$N\rightarrow\infty$ \cite{LANDI} required by M(atrix) theory corresponds,
in the dual type IIB description, to considering the ${\cal K}$--bundles
of \cite{BOUWMATHAI}, rather than the Azumaya algebras of \cite{KAPUSTIN}.

Some comments are in order. Assume appling $p-1$ T--dualities 
instead of $p$, to get a system of D1--branes. The D1--brane is S--dual to the  
fundamental type IIB string. The latter can be wrapped on $\Sigma$  
at the cost of breaking  all supersymmetry \cite{RUSSO}. Hence the  
D$p$--brane wrapped on $\Sigma$ breaks all supersymmetry,  too, and it  
corresponds to a non--BPS configuration.

The D1--brane can be viewed as the strong--coupling limit 
of the fundamental type IIB string in 10 dimensions. On the other hand, 
11--dimensional M(atrix) theory is a model for M--theory, {\it i.e.}, 
for the strong--coupling limit of type IIA string theory. Moreover, 
T--duality being a perturbative symmetry, it will not exchange the weak 
and the strong--coupling regimes. This accounts for the mismatch 
of dimensions between the two dual descriptions we have given.  
  
\subsection{The limit $N\rightarrow\infty$}\label{limit}  

By equation (\ref{totalanomaly}) we have specified a class $[H]_Q$. 
In the limit $N\rightarrow\infty$, this $[H]_Q$ specifies 
an isomorphism class of ${\cal K}$--bundles over $Q$. 
Picking a torsion class in $H^3(Q, {\bf Z})$ amounts to picking an 
isomorphism class of projectively flat bundles ${\cal E}\rightarrow Q$
with fibre ${\cal K}$ and structure group $PU({\cal H})$. If we now choose 
the manifold $Y$ as explained in section \ref{manifoldy} below, 
then such an isomorphism class of bundles is specified by a representation 
of $\pi_1(\Sigma)$ into $PU({\cal H})$.

As summarised in section \ref{defcstar}, in \cite{PROCEEDINGS} we have 
explicitly constructed, on the separable  Hilbert space ${\cal H}=L^2({\bf H})$, 
a 1--parameter family  $\rho_b$, $b\in {\bf R}$, of projectively unitary representations 
of the Fuchsian group $\Gamma\simeq\pi_1(\Sigma)$ uniformising $\Sigma$.
Although infinite--dimensional, these representations $\rho_b$ can be understood 
as the double--scaling limit $M\rightarrow -\infty$,  $N\rightarrow\infty$, 
of the Narasimhan--Seshadri representations $\rho_{NM}$ reviewed in section \ref{nstheorema}. 
The latter represent $\pi_1(\Sigma)$ on ${\bf C}^N$, where $N$ is the rank 
of the gauge group $U(N)$ carried by the stack of $N$ coincident branes, 
and $M\in{\bf Z}$ is the 't Hooft magnetic flux obtained integrating 
the 't Hooft 2--form $[f]$ over $\Sigma$. The parameter $b\in{\bf R}$ 
on which $\rho_b$ depends can be fine--tuned at will. The identification between  
our $\rho_b$ of equation (\ref{inoxnni}), and its finite--dimensional counterpart 
$\rho_{NM}$ of Narasimhan--Seshadri,  equation (\ref{nsmatrices}), 
proceeds as follows. The $N\times N$ unitary matrices $u_j$ acting 
on ${\bf C}^N$ become unitary operators ${\cal U}_j$ acting on $L^2({\bf H})$,
\begin{equation}  
\lim_{N\to\infty,\,M\to -\infty}u_j={\cal U}_j,
\label{identification}
\end{equation} 
and the phase multiplying the identity on the right--hand side of 
(\ref{nsmatrices})
is identified with that on the right--hand side of (\ref{inoxnni}),
\begin{equation}
\lim_{N\to\infty,\,M\to -\infty}
\exp \left(2\pi i {M\over N}\right)=
\exp \left(-2\pi i \theta_b\right).  
\label{identificationx}
\end{equation} 
In this way we have determined a 1--parameter family of projectively flat 
${\cal K}$--bundles ${\cal E}_b\rightarrow\Sigma$. We conclude that   
our infinite--dimensional representations $\rho_b$ of $\pi_1(\Sigma)$
of equation (\ref{inoxnni}) are induced by turning on a 't Hooft magnetic 
flux across the Riemann surface $\Sigma$ inside the worldvolume of the 
$N=\infty$ coincident D$p$--branes. 

As we have seen in section \ref{defcstar}, one can interpret the infinite--dimensional 
representation of $\pi_1(\Sigma)$ given in \cite{PROCEEDINGS} as defining a
noncommutative $C^{\star}$--algebra $C^{\star}(\Gamma, \theta)$. Through the 
Sen--Witten construction, the latter is the result of turning on a nonzero 't Hooft 
magnetic flux in the worldvolume of the $N'=\infty$ D$p'$--brane/antibrane pairs 
that are equivalent to $N=\infty$ coincident D$p$--branes wrapped on $\Sigma$. 
Alternatively, through the anomaly cancellation condition, this flux is due to 
turning on a background Neveu--Schwarz $B$--field.  

\subsection{Choice of the fibre bundle over $\Sigma$}\label{manifoldy}

Given that $H^3(\Sigma, {\bf Z})$ is trivial, one would like to wrap the D$p$--branes 
on a manifold whose real dimension is greater than 2. This would allow the
correspondence between ${\cal K}$--bundles and classes in $H^3(\Sigma, {\bf Z})$
a possibility of being nontrivial. Fibre bundles over the Riemann surface $\Sigma$ thus come 
to mind. We will not attempt a complete classification of all 
possibilities, as in fact trivial bundles over $\Sigma$ will suffice.
We will satisfy ourselves with an example of a trivial bundle
$\Sigma\times Y$, for a certain choice of the spacetime manifold $Y$,
that will allow for a nontrivial torsion. Again it will turn out that more 
than one choice for $Y$ is possible. The spacetime manifold $Y$ will be determined 
imposing consistency conditions. 

In type IIB superstring theory, the manifold $Y$ must be orientable
and spin. Furthermore, $Q=\Sigma \times Y$ must allow for a nontrivial 
torsion subgroup within $H^3(Q, {\bf Z})$. Finally, $H^3(Q, {\bf Z})$ 
parametrises isomorphism classes of ${\cal K}$--bundles over $Q$, 
but instead we need it to parametrise isomorphism classes of 
${\cal K}$--bundles over $\Sigma$. Hence $Y$ must be chosen 
in such a way that torsion classes in $H^3(Q, {\bf Z})$ 
continue to parametrise ${\cal K}$--bundles over $\Sigma$.

This refines the minimum value of $p$ determined in section \ref{ann}, 
where it was found that $p\geq 3$. A nontrivial $H^3(Q, {\bf Z})$ further 
imposes $p>3$. Indeed, $p=3$ would correspond to a $1+1$ dimensional $Y$.
Factorise it (at least locally) as the product of a timelike factor $Y_t$ 
times a spacelike factor $Y_x$. The latter can be chosen compact 
or not, which leads to these topologically different choices for $Y_x$: 
$S^1$ and ${\bf R}$, and quotients thereof, such as ${\bf RP}^1$, for 
example. One finds that none of these choices satisfies our needs.
Taking $Y_x={\bf R}$ leads to a trivial $H^3(Q, {\bf Z})$.
The choice $Y_x=S^1$, while producing a nontrivial $H^3(Q, {\bf Z})$, 
is torsionless; so is the case of ${\bf RP}^1$.

Within the type IIB theory the next allowed value is $p=5$. 
Again separating out the trivial timelike dimension, let us see 
that one can find a spacelike manifold $Y$ in real dimension 3
satisfying the necessary requirements.

For the correspondence between torsion classes in $H^3(Q, {\bf Z})$
and ${\cal K}$--bundles over $\Sigma$ to hold, one would on first sight
require $Y$ to have a trivial fundamental group, so that $\pi_1(Q)=\pi_1(\Sigma)$. 
However, this condition can be relaxed to a less stringent one. 
We will see presently that an abelian $\pi_1(Y)$ will suffice.
Kunneth's formula \cite{BOTT} allows us to write
\begin{eqnarray}
H^3(\Sigma\times Y, {\bf Z}) \subset
H^0(\Sigma, {\bf Z})\otimes H^3({\bf Y}, {\bf Z}) \oplus 
H^1(\Sigma, {\bf Z})\otimes H^2({\bf Y}, {\bf Z}) \cr
\oplus
H^2(\Sigma, {\bf Z})\otimes H^1({\bf Y}, {\bf Z}) \oplus 
H^3(\Sigma, {\bf Z})\otimes H^0({\bf Y}, {\bf Z}).
\label{decomp}
\end{eqnarray}
In the particular case at hand, one can show that the above inclusion is 
actually an equality.
Now $H^3(\Sigma, {\bf Z})$ is identically zero, while
$H^0(\Sigma, {\bf Z})={\bf Z}=H^2(\Sigma, {\bf Z})$ and $H^1(\Sigma, {\bf Z})=
{\bf Z}^{2g}$. Torsion pieces, if any, must come from $H^3(Y, {\bf Z})$, 
$H^2(Y, {\bf Z})$ and $H^1(Y, {\bf Z})$. Allowing for an abelian $\pi_1(Y)$ 
for the moment, the manifold ${\bf RP}^3$ (which is orientable and spin)
has a nontrivial torsion 
\begin{equation}
H^1({\bf RP}^3, {\bf Z})={\bf Z}_2.
\label{sitors}
\end{equation}
More generally, branes on group manifolds have been studied in \cite{FS}.

It remains to explain why one can allow for an abelian $\pi_1(Y)$ without
spoiling the 1--to--1 correspondence between torsion classes in $H^3(Q, 
{\bf Z})$ and isomorphism classes of ${\cal K}$--bundles over $\Sigma$. 
The latter are in bijective correspondence with projectively unitary 
representations of $\pi_1(\Sigma)$. Now the decomposition 
$\pi_1(Q)=\pi_1(\Sigma)\times\pi_1(Y)$ together with equation 
(\ref{inoxnni}) provides the answer: factors coming from an abelian $\pi_1(Y)$
will cancel when computing the left--hand side of (\ref{inoxnni}). (We could 
even allow for a projectively represented abelian group $\pi_1(Y)$, at the 
cost of considering its nontrivial contribution to right--hand side of 
(\ref{inoxnni})).

We close this section with an observation. The anomaly--cancellation 
condition is key to our construction. We have applied it within type 
IIB superstring theory, in oder to link it to the Sen--Witten 
superposition of branes with antibranes. However, one could just as well 
apply it to bosonic string theory, where nonorientable manifolds
are allowed and the anomaly--cancellation condition \cite{KAPUSTIN} 
simplifies to 
\begin{equation}  
\beta_N([f])=[H]_Q.  
\label{bosonicanomaly}
\end{equation}  
The requirements on the manifold $Y$ thus become less restringent,
and one can verify that the following examples satisfy all our needs.
The 2--dimensional real projective space ${\bf RP}^2$  and the Klein
surface ${\bf K}^2$ have nontrivial torsion given by
\begin{equation}
H^1({\bf RP}^2, {\bf Z})={\bf Z}_2, \qquad H^1({\bf K}^2, {\bf Z})={\bf 
Z}\oplus {\bf Z}_2.
\label{nontors}
\end{equation}
The absence of supersymmetry in our construction (see also section 
\ref{morgplus}) allows us to consider these possibilites as valid for 
the physical realisation of $C^{\star}(\Gamma, \theta)$ in terms of 
strings and branes.

\section{BPS--like spectra in $g>1$ from the Narasimhan--Seshadri theorem}
\label{morgplus}
  
In $g=1$, Morita equivalence of noncommutative gauge theories is reflected 
in the T--duality of superstring theory \cite{PIOLINE}. If we were to follow 
the reasoning applied in $g=1$ \cite{SCHWARZ}, we would now have to identify 
the dual tessellation $T^*({\bf H})$. The latter would parametrise the  
endomorphisms ${\rm End}\,E_{NM}$ of the module $E_{NM}$. However, 
$T^*({\bf H})$ must be a quantum space, since $\Gamma$ is nonabelian.
Moreover, in $g>1$ there is no T--duality, and compactification breaks 
all supersymmetry \cite{RUSSO}. Hence, unlike in $g=1$, there are no 
supersymmetric BPS spectra in $g>1$.  This notwithstanding, the breakdown 
of supersymmetry does not prevent the existence of stable, non--BPS states 
in M--theory \cite{WITTENDK, SEN}.  
 
We will therefore follow an alternative route. We will prove the existence 
of constant--curvature connections on the projective modules $E_{NM}$. We 
will see that, as in $g=1$, in $g>1$ there exists a moduli space of 
such connections. Even though there is no supersymmetry, one can take such 
connections as defining the $g>1$ analogues of BPS states on the torus, 
since the latter were also characterised as having constant curvature.
In $g=1$ the stability of such states was a consequence of supersymmetry.
In the absence of supersymmetry, however, the stability of these states deserves 
a separate study. 

\subsection{Constant--curvature connections on $E_{NM}$}\label{cccon}  
  
The finite--dimensional space ${\bf C}^N_{(M)}$ in equation (\ref{totspace})
is the fibre of a stable holomorphic bundle over $\Sigma$. Let us 
assume that the double--scaling limit $M\to -\infty$, $N\to\infty$ 
respects stability. In other words, we assume that this limit can 
be taken in such a way that $L^2({\bf H})$ becomes the fibre of an 
(infinite--dimensional) stable holomorphic bundle over $\Sigma$.
Then a suitable infinite--dimensional generalisation of Donaldson's version 
of the Narasimhan--Seshadri theorem establishes the existence of a metric 
connection $\nabla_D$ such that the constant--curvature condition (\ref{central}) 
\begin{equation}  
F_D=-2\pi i \left({M\over N}-\theta_b\right)\,\omega\,{\bf  I} 
\label{constanze}
\end{equation}  
holds. Above, $\omega$ equals the Poincar\'e 2--form 
${\rm d}z\wedge {\rm d}\bar z/({\rm Im}\,z)^2$ on ${\bf H}$, 
and ${\bf I}$ denotes the identity on $E_{NM}$.

A remark is in order. There is a formal analogy between the last equation 
in (\ref{konsch}) and equation (\ref{constanze}). However, contrary to the 
noncommutative torus, our noncommutative $C^{\star}$--algebra $C^{\star}(\Gamma,\theta)$ 
and its projective modules cannot be obtained from the representation theory 
of the Heisenberg algebra. In fact we have followed a route different 
from that of the noncommutative torus \cite{SCHWARZ}. In $g=1$ one first 
constructs a derivation $\delta$ of the $C^{\star}$--algebra. 
Next one uses $\delta$ in order to define a connection $\nabla$. 
Finally $\nabla$ is used, as in equation (\ref{konsch}), in order to 
impose the constant--curvature condition. In $g>1$ we have bypassed this
procedure because the constant--curvature condition (\ref{constanze}) 
is no longer a Heisenberg algebra. Without defining a derivation $\delta$
of $C^{\star}(\Gamma, \theta)$, the Narasimhan--Seshadri theorem directly 
allows us to construct the desired connections on the projective modules 
$E_{NM}$.

\subsection{Moduli space of constant--curvature connections}\label{xyz}  
  
The previous construction relied on the notion of stability for 
holomorphic vector bundles over $\Sigma$. As we have seen, stability is 
required in order to have constant--curvature connections or, in physical 
terms, BPS--like states. There is one more reason to require stability.
In $g=1$ there exists a moduli space of BPS states. Does a moduli space 
of BPS--like states exist in $g>1$?
  
Topological vector bundles over $\Sigma$ are classified, up to isomorphism, by  
the rank $N$ and the degree $M$. However, the classification of {\it  
holomorphic} vector bundles involves continuous parameters, so we have a  
moduli space of holomorphic vector bundles over $\Sigma$. From the above it  
follows that this moduli space coincides with that of constant--curvature  
connections. The latter define the higher--genus analogue of BPS states. So the  
$g>1$ analogues of BPS states are parametrised by the points of the moduli space  
of holomorphic vector bundles. It turns out that the latter space in general  
is not Hausdorff, but the condition of stability suffices to ensure a good  
moduli space. The precise statement is as follows \cite{THADDEUS}:  
fix the data $\Sigma$, $N$ and $M$, the latter two coprime. Then there  
exists a complex smooth, connected and compact moduli space  
${\cal M}^{(g)}_{NM}$ of  
equivalence classes of rank $N$, degree $M$, stable holomorphic vector bundles  
over $\Sigma$, with dimension $N^2(g-1)+1$. The moduli space  
${\cal M}^{(g)}_{NM}$ depends only on the residue class of $M$ modulo $N$. 

\section{Conclusions and outlook}\label{outlook} 
 
In this paper we have established an interesting link between 
noncommutative geometry and the Sen--Witten construction of non--BPS branes,
by explicitly constructing a noncommutative $C^{\star}$--algebra 
$C^{\star}(\Gamma, \theta)$ that generalises to $g>1$ what the 
noncommutative torus does in $g=1$. The mathematical definition of 
$C^{\star}(\Gamma, \theta)$ was presented 
in \cite{PROCEEDINGS}; in this paper it has been given a physical interpretation 
in terms of the wrapping of D$p$--branes on a Riemann surface $\Sigma$ in $g>1$, 
with a background $B$--field turned on. The latter deforms the commutative 
$C^{\star}$--algebra of functions to a noncommutative $C^{\star}$--algebra 
that we have succeeded in identifying. Finally, we have constructed 
a family of projective modules over $C^{\star}(\Gamma, \theta)$ 
and proved the existence of constant--curvature connections on them.

In $g=1$, Morita equivalence led to a whole ${\rm SL}_2({\bf Z})$ orbit of  
Morita--equivalent noncommutative tori \cite{SCHWARZ}. This was due to the 
Abelian property of the fundamental group of the torus, which allowed for 
an easy identification of the commutant. However, the fact that the Fuchsian 
group uniformising a Riemann surface in $g>1$ is nonabelian implies that 
there exists no Morita--group orbit of $C^{\star}(\tilde\Gamma, \tilde\theta)$ 
algebras that are Morita--equivalent to $C^{\star}(\Gamma,\theta)$. 
This notwithstanding, we have succeeded in identifying the $g>1$ analogues 
of supersymmetric BPS states on the noncommutative torus, 
thanks to Donaldson's description of stable vector bundles 
over Riemann surfaces.

An important physical question to address in this context is the 
stability of the BPS--like states constructed here. It would be very 
interesting to relate the mathematical property of stability of holomorphic 
vector bundles with the physical property of being a stable, non--BPS state.
Mathematically, one would like to compute the topological numbers 
and the Chern character for the projective $C^{\star}$--modules $E_{NM}$. 
We hope to report on these issues in the future.
 
{\bf Acknowledgments} 

This work has been supported by a PPARC fellowship under grant no. PPA/G/O/2000/00469.

\end{document}